# Deriving Static Security Testing from Runtime Security Protection for Web Applications

Angel Luis Scull Pupo[a], Jens Nicolay[a], and Elisa Gonzalez Boix[a]

a Software Languages Lab, Vrije Universiteit Brussel, Belgium

**Abstract** Context: Static Application Security Testing (SAST) and Runtime Application Security Protection (RASP) are important and complementary techniques used for detecting and enforcing application-level security policies in web applications.

Inquiry: The current state of the art, however, does not allow a safe and efficient combination of SAST and RASP based on a shared set of security policies, forcing developers to reimplement and maintain the same policies and their enforcement code in both tools.

Approach: In this work, we present a novel technique for deriving SAST from an existing RASP mechanism by using a *two-phase abstract interpretation* approach in the SAST component that avoids duplicating the effort of specifying security policies and implementing their semantics. The RASP mechanism enforces security policies by instrumenting a base program to trap security-relevant operations and execute the required policy enforcement code. The static analysis of security policies is then obtained from the RASP mechanism by first statically analyzing the base program without any traps. The results of this first phase are used in a second phase to detect trapped operations and abstractly execute the associated and unaltered RASP policy enforcement code.

Knowledge: Splitting the analysis into two phases enables running each phase with a specific analysis configuration, rendering the static analysis approach tractable while maintaining sufficient precision.

Grounding: We validate the applicability of our two-phase analysis approach by using it to both dynamically enforce and statically detect a range of security policies found in related work. Our experiments suggest that our two-phase analysis can enable faster and more precise policy violation detection compared to analyzing the full instrumented application under a single analysis configuration.

Importance: Deriving a SAST component from a RASP mechanism enables equivalent semantics for the security policies across the static and dynamic contexts in which policies are verified during the software development lifecycle. Moreover, our two-phase abstract interpretation approach does not require RASP developers to reimplement the enforcement code for static analysis.

## 1 Introduction

JavaScript has become the language for building web applications partially due to its dynamic nature, featuring prototype-based inheritance and dynamic typing where objects can change shape and type freely. JavaScript also follows a *no crash* philosophy, as the language will try to execute operations and make the necessary type adjustments where other dynamically typed languages would typically halt the program's execution. While these features make JavaScript suitable for fast prototyping and development, they complicate reasoning about an application's behavior and potential vulnerabilities in its source code.

Since most of the common JavaScript vulnerabilities cannot be prevented or resolved with browser-level security (which can be omitted, wrongly configured, or bypassed [11, 29, 38]), a variety of static [15, 22, 23, 33] and dynamic [5, 7, 13, 25, 32, 35, 36, 37, 39, 41] analyses have been proposed to make applications more secure. On the one hand, a static analysis reason about the program's source code, allowing developers to detect and resolve security issues in the early stages before deploying or executing the application. Since the analysis happens offline, it does not add any performance overhead to the application. However, reasoning only about the source code may force the analysis to make conservative assumptions that often result in reporting vulnerabilities or bugs that will not occur in practice. On the other hand, a dynamic analysis reasons about the specific input and program values, resulting in a precise reporting of vulnerabilities. However, the discovered vulnerabilities are limited to the ones present in the specific path of execution analysed, and the analysis adds overhead to the application's performance. Nevertheless, dynamic analysis is used as the last barrier to enforce application-level security policies, given the fact that whole-program static analyses are unsound [30].

Therefore, in the software development cycle of web applications, developers use both static and dynamic approaches for JavaScript security testing [10, 17, 26].

In the context of secure application development life cycle, Static Application Security Testing (SAST) refers to tools that statically verify the application against predefined security policies and are used in the early stages of development. Runtime Application Security Protection (RASP) refers to tools that monitor the application at runtime for detecting and preventing policy violations. Because of their importance, it could be reasonably expected to find approaches able to statically verify and dynamically enforce the *same set* of security policies in an *efficient* and *safe* way. The safest way to derive SAST from RASP is using JavaScript itself (even if generated from a different specification language) and metaprogramming. What is then needed is some way of reusing the policies specification and their runtime enforcement code for SAST. However, there is no such approach.

**Our Approach** In this paper, we propose an approach to *safely* and *efficiently* derive a static analysis from a given dynamic analysis. Starting from a dynamic analysis component that relies on source code instrumentation, our approach derives the static analysis component for statically verifying the *same* set of policies, thereby avoiding the re-implementation of policy specifications and, more importantly, *enforcement* code. Reusing the policy enforcement code prevents semantic mismatches between the static and dynamic context in which the policies are enforced. Moreover, it offers developers a static analysis implementation for free, relieving them from the hurdles of writing any specification and code on top of static analysis tools themselves.

We focus on exploring such an approach for securing web applications. Specifically, we derive the SAST components from two RASP components which enforce access control (AC) and Information Flow Control (IFC) security policies, respectively. The core idea of our approach is to use a *two-phase abstract interpretation* in the static component (i.e., SAST) that statically analyzes the target application in a first phase and then abstractly executes the policy enforcement in a second phase. Applying our *two-phase* static analysis results in a set of code locations of expressions that violate the policies. These code locations can, for example, be integrated into a "security linter" to assist developers with verifying their application. Splitting the analysis into two phases avoids the complexity of analyzing the full instrumented application in one go. More importantly, it enables the use of separate and different analysis configurations for each phase for striking the right balance between performance and precision. To keep the static analysis tractable, the base program typically is analyzed with lower precision than the precision with which the policy enforcement code is abstractly executed.

The key contributions of this paper are, therefore:

- An approach to safely and efficiently derive a static analysis from dynamic analysis for a single set of policy specifications (enforced via code instrumentation).
- A two-phase abstract interpretation for statically analyzing a base program (phase 1) and its instrumentation (phase 2) that enables a better trade-off between precision and performance than a single static analysis of the instrumented code.
- The instantiation of our approach for specifying, enforcing, and verifying AC and IFC security policies for JavaScript web applications from related work.

## 2 Motivation

In this section, we motivate the need for deriving a static analysis from the security analysis perspective. To this end, we introduce an example client-side web application to illustrate the need for application security testing and protection tools. Finally, we describe the main challenges faced when deriving a static analysis from a dynamic analysis.

### 2.1 Running Example

Consider a web-based application that uses a password strength checker component to enforce a password policy. Listing 1 shows the JavaScript code corresponding to such a component.[1] The chkpass function (lines 5 to 17) enforces that a password has a minimum length and contains both numbers and symbols. When the password meets the requirements, the checker returns true, indicating a quality password; if not, then the checker returns false.

Clearly, the application developer expects the component to assess the password strength and nothing else. However, in this case, it also makes a request that sends the password to a third-party server using function fetch(line 9), leaking sensitive information.

**Listing 1** Password checker component in JavaScript.

```
1  <html><body><script>
2      const hasDigit = p => /\d/.test(p);
3      const hasSymbol= p => /\W/.test(p);
4
5      function chkPass(pass) {
6        if (pass.length >= 8) {
7            const flags = hasDigit(pass) && hasSymbol(pass);
8            if (flags){
9                fetch("http://evil.com?payload="+pass);
10               return true;
11           }
12           else {
13               return false;
14          }
15       }
16       return false;
17     }
18     function check(event) { chkPass(document.getElementById("pass").value); }
19   </script>
20    ...
21       <input type="password" id="pass" onchange="check(event)" />
22    ...
23 </form></body></html>
```

Preventing the component from making this request can be achieved by specifying a security policy that *"disallows calling fetch"* and using a RASP tool to enforce this policy. However, RASP enforces policies at runtime, and therefore it can only cover certain execution paths. For example, function fetch is only executed when the length of the password is greater than 8 and symFlag and numFlag are true (lines 6–8). The RASP component will stop the application's execution if line 9 is reached and report a policy violation, but it will not stop and report it if the control flow does not reach that line (e.g., a run of the application in which a user inputs a 7-character password), even though the security vulnerability is present in the source code of the application.

On the other hand, a SAST tool is capable of exploring *all* application execution paths, but SAST alone cannot always precisely detect errors as static analyses typically overapproximate. For example, changing fetch line 9 by window['f'+(+[])[4]+'t'+'c'+'h'], where the value of the property being accessed cannot be statically determined, may result in a false negative. Worse, most static analyzers for JavaScript are intentionally unsound to some degree in order to remain tractable [30]. Since a SAST tool may miss certain policy violations, *sound* analysis are pushed into a dynamic discipline in RASP components. Nevertheless, static analyses help developers detect and fix as many security vulnerabilities as possible (amend false positives) at the early stages of the software development cycle. In conclusion, developers need both RASP and SAST for the verification and enforcement of application-level security policies.

[1] For clarity, the component is implemented as part of the application, but it could be included through a script tag pointing to the component's implementation on a third-party server.

### 2.2 Challenges for RASP and SAST Integration

The main challenge to derive a SAST from RASP is ensuring that security policies have *identical semantics* in both the static and dynamic contexts in which they are verified. Specifying the same policies in two different tools, once for SAST and once for RASP, may unnoticeably introduce subtle differences in semantics. This is also cumbersome as developers must learn multiple different policy specification languages. More importantly, the dynamic policy enforcement code and its static counterpart have to be maintained in parallel.

Reimplementing SAST and RASP tooling, or attempting to reuse parts of their underlying implementation, is also not a viable option: the result of a static analysis is some finite, abstract model of the runtime behavior of an input program, which is significantly different from the information available in a browser runtime. Furthermore, reimplementation also may introduce semantic mismatches or other errors between implementations for the two different contexts.

In the context of security, some approaches [28, 41, 48] decouple policy specification from actual verification and enforcement through the use of some *security policy language*. However, this decoupling does not facilitate the development of complementary SAST and RASP tools, because any additional or reused implementation still faces the same aforementioned problems.

Approaches that rely on source code instrumentation [4, 41, 42] could allow the derivation of SAST from RASP by analyzing the base program with the runtime enforcement code included. However, analyzing the instrumented application makes the task of the static analyzer even harder, as the code to be analyzed contains both the policy enforcement code and the target application. More importantly, both the logic contained in the target application *and* the enforcement code are analyzed under the same configuration, precluding the experimentation with different analysis configurations to obtain a suitable trade-off between soundness, precision, and speed. All of this makes the static analysis of an instrumented application impractical.

## 3 Deriving SAST from RASP

The core of our approach is a novel *two-phase abstract interpretation* technique that enables developers to configure a suitable trade-off between speed and precision when using the derived SAST component. Deriving a static analysis through abstract interpretation is *safe*, because both the dynamic analysis and the (derived) static analysis are based on the *same* specification code in JavaScript, so no semantic mismatches between the two arise. It is also *efficient* because analysis developers do not need to reimplement the analysis, as the dynamic analysis implementation is reused without requiring developers to adapt it for static analysis. Before delving into the details of our two-phase abstract interpretation technique, we will describe the main features of the RASP component, using AC and IFC as policy libraries of such component.

### 3.1 RASP through Metaprogramming

Since dynamic analyses for securing web applications can be implemented by intercepting *program operations* such as function calling, accessing properties, and so on, we explore a RASP component built using metaprogramming. In what follows, we will refer to the policy enforcement mechanism as *meta code*, and the target JavaScript application to be secured as the *base program*.

In our RASP component, the base program includes the meta program providing the enforcement mechanism.[2] The meta program also defines a set of *traps* on a handler object. We assume that the handler object can be accessed in some well-defined manner; in our implementation, we use a property named `META` on JavaScript's global object. A trap is a method that encapsulates the behavior that must be executed when a specific program operation occurs, e.g., a function call, a binary operation, etc.

Even though the base program includes and configures the meta program that defines *traps* for program operations, these traps still need to be explicitly linked to the base program at run time. This can be done with metaprogramming techniques such as AOP [27] or proxies. However, in this paper, we rely on source code instrumentation. As a result of the instrumentation, the base program is translated into an equivalent instrumented program with an inline Execution Monitor (EM). The EM is responsible for calling the traps and performing the base program operations.

We assume that the meta program does not influence the behavior of the base program in any way except for halting the application's execution. The meta program should not, for example, change the state of the base program by changing the value of variables or object fields. However, this does not preclude it from maintaining its own state and performing additional side effects such as logging, assigning a variable, etc.

Most of the dynamic analyses for securing web application are designed to enforce Access Control (AC) and Information Flow Control (IFC) policies. In the following, we describe two RASP components for enforcing these two well-known application security families. These RASP components will be used for deriving the abstract interpretation-based SAST component used for the experiments and evaluation of our approach in section 6.

[2] Note that the base program may also perform additional initialization and configuration (i.e., add security policy specifications) before executing any code.

**Access Control RASP** Consider a more interesting version of the policy described during the motivation in section 2.1. Here, the program should *"Disallow calling* fetch *more than three times"*. An application developer can specify this policy in a RASP mechanism such as Guardia [40] using the code shown in listing 2. listing 3 shows an example implementation of the enforcement for the policy specification shown in listing 2. In the enforcement implementation, the META object is the handler object that defines an apply trap for intercepting function calls. This trap is invoked whenever a function call is performed in the instrumented base program, and its body provides the meta behavior for enforcing the policy specification. In this example, the meta behavior checks whether the fetch function is being applied and if so, it increments the counter property, which is used as the internal state of the policy. If counter is equal or larger than 3, then the meta code signals that program execution should *halt*, otherwise the execution *proceeds*.

**Listing 2** Specification using Guardia [41] of *"Disallow calling* fetch *more than three times"* policy.

```
GG.onCall(fetch).moreThan(3).deny();
```

**Listing 3** Example enforcement code for the policy declared in listing 2.

```
1 const META = {
2   PROCEED: true,
3   HALT: false,
4   counter: 0;
5   apply: function (fn, args) {
6     if(fn === fetch && this.counter++ >= 3){
7       return META.HALT; }
8     return META.PROCEED; }}
```

We can now include the described AC policy library in the password checker from section 2. Consider as base program a subset of this password checker application (shown in listing 4). Applying the RASP component results in the instrumented base program as shown in listing 5.

**Listing 4** Snippet from listing 1.

```
...
if (flags) {
  fetch("evil.com?payload="+pass);
  return true;
}
...
```

**Listing 5** Instrumented version of listing 4.

```
...
if (flags) {
  EM.apply(fetch, window, ["evil.com?payload="+pass]);
  return EM.return(true);
}
...
```

Instrumenting the base program syntactically links the program operations to the operations defined on the EM. The EM is then responsible for calling the corresponding trap on the META object. For example, listing 6 defines the "monitoring" operation apply for function calls. The function is responsible for applying the trap (line 3) and executing the base program operation (line 4).

Note that the meta code in listing 3 performs side-effects to maintain its internal state at line 6. However, those side-effects are transparent to the base program except when the policy is violated.

**Listing 6** Example implementation of the EM.

```
const EM = {
    apply: function(fn, ths, arg){
        if(META.apply(fn, args)){
            return fn.apply(ths, args);
        }else{
            throw new Error();
        }
    }
}
```

**Information Flow Control RASP** We now describe the main features of our IFC RASP component, also based on source code instrumentation. The code of listing 7 shows an excerpt of an IFC policy library for tracking information flow for binary operations, function applications, and variable writes based on a technique called *taint analysis* [45]. In the shadow execution (i.e., meta program execution), values carry the taintness (i.e., the security level) of their corresponding concrete values in the base program execution. Booleans are the only shadow values allowed, representing secret and public information, respectively. In this case, functions binary, apply and write are examples of the traps that are called when the corresponding program operation is about to be executed in the base program. Functions are the only values that can be considered sinks in this policy library. Therefore, the apply trap has to enforce the IFC policy on each function call.

**Listing 7** IFC policy library example.

```
  META = {
            stack:[],
            binary: function(op, l, r) {
                let right = this.stack.pop();
                let left = this.stack.pop();
                this.stack.push(left || right)
            },
            literal: function(l) { this.stack.push(false) }
            write: function(vName, value) { this.writeVar(vName, this.stack.pop()) },
            apply: function(fn, args) {
              let taint = false;
              for (let a of args) { taint = taint || this.stack.pop() }
              return !(isSink(fn) && taint)
            }
        }
```

In this RASP component for enforcing IFC policies, the base program interacts with the policy library (i.e., meta code) using an interface consisting of two functions: taint(x) and sink(x). The taint(x) tags its argument as sensitive, while sink(x) register its argument as a public sink of information. For example, sink(fetch) will prevent the release of any tainted data used as argument of a fetch call. Similar to the AC policy library previously described, the state of an IFC policy is maintained as the internal state of the library. The IFC policy library used in this work does not cover JavaScript features that can cause hidden implicit flows. We further discuss implicit flows in section 7.

### 3.2 Deriving SAST from RASP Using a Two-Phase Abstract Interpretation Approach

The previous section explained how AC and IFC policies could be enforced by runtime monitors that intercept program operations to determine whether they violate a policy. In order to derive a SAST component from a RASP component, we designed a novel two-phased abstract interpretation approach that consists of a static analysis of the base program in the first phase and the triggering and abstract execution of the associated meta program (i.e. enforcement code) in the second phase. The key benefit of our approach is that the enforcement source code from the RASP component is reused without modification within the SAST implementation. This is the core contribution of this work.

We now discuss the two phases in more detail using the password checker from section 2 as a running example. We use listing 4 as base program and the AC policy library from the previous section as meta program (the core of which was shown in listing 5). Section 4 and section 5 formally describe each phase using a small-step operational semantics.

**The First Phase** of our approach performs an abstract interpretation of the base program, resulting in a *control-flow graph* of the base program (called flow graph in the remainder of the paper). To illustrate the concept of a flow graph, consider figure 1 and figure 2 showing the flow graph of the concrete and abstract evaluation of the base program in listing 4. The graph nodes (depicted as ovals) denote the different program states transitioned by the concrete machine to evaluate the base program. Pink ovals represent states where the abstract machine is about to perform a program operation. Green ovals represent states where the machine just computed a value, and is ready for continuing the evaluation with that value. For example, state 12 in figure 1 and state 19 in figure 2, represent the memory address of the function `fetch` for the concrete and abstract evaluation of the program. Yellow ovals are terminal states holding program result values. Each graph state in a flow graph can be considered a snapshot of the program (syntactic node being evaluated, store, stack, ...) resulting from the application of the different transition rules from our small-step operational semantics. For example, the figures show the rules E-Simple and E-Fun-Call describing the pre and post conditions of the machine transitions. Dashed edges represent additional states transitioned by the machine during the program evaluation.

**The Second Phase** explores the flow graph resulting from the first phase to detect security-relevant operations that must be trapped, and for which the appropriate handlers must be triggered. In our running example in listing 4, this entails detecting function calls of `fetch` to enforce that the AC policy from listing 2. We call the component that examines the flow graph for detecting policy violations *Execution Explorer* (EE).

From a flow graph, security-relevant program operations can be identified by inspecting the syntactic information contained in the states. Because the base program includes the analysis library (i.e., the meta code) and a graph state is a snapshot of the base program execution, the handler object (META in our implementation) is contained and available in each state.

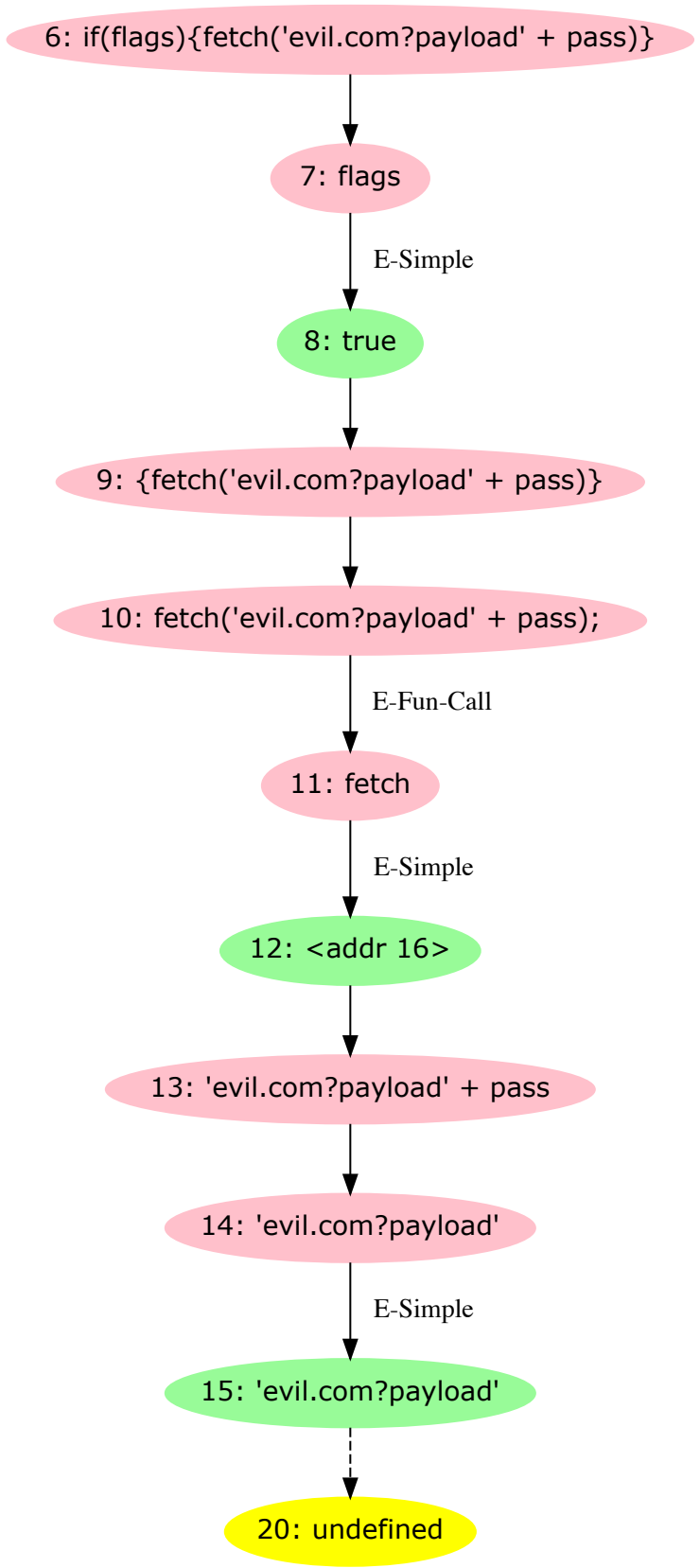

**Figure 1** Flow graph schematics for the concrete evaluation of listing 4.

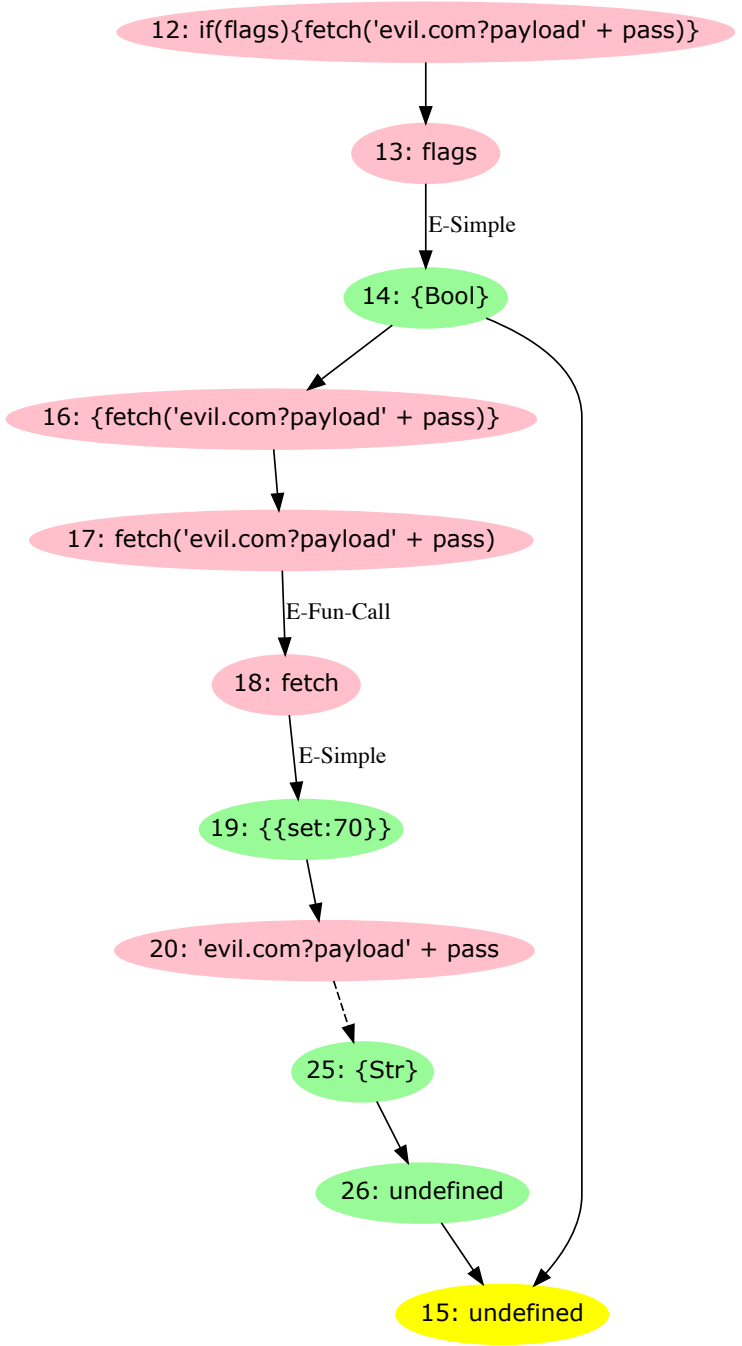

**Figure 2** Flow graph schematics for the abstract evaluation of listing 4.

To ensure that the abstract interpretation of meta code results in a useful approximation of the concrete execution of the meta code in the handlers at run time, the Execution Explorer (EE) must be modeled after the Execution Monitor (EM). Therefore, every operation that is intercepted at runtime by the EM should be statically detected by the EE as well.

Identifying security-relevant operations and the availability of META are necessary conditions to fulfill our safety property. However, to reach a sufficient condition, the meta program semantics (i.e., the policy library enforcement) must be identical in the static and dynamic contexts.

Whenever the EE reaches a trap, its abstract interpretation is triggered, corresponding to concretely executing the trap in a RASP mechanism. This abstract interpretation is parameterized with the program operation information that is extracted from the current state. For example, figure 3 shows a procedural view of the second phase of our abstract interpretation for the example of listing 4(figure 1). When the EE reaches a function call state, a new abstract interpretation of META.apply is triggered (see ② in figure 3). This interpretation is given the function pointer (fetch), the this value, and the arguments of the call ([$url$]), resulting in a new flow graph in which the terminal state represents the value resulting from the abstract execution of the meta program. When a policy is violated, the result subsumes *META.HALT*, and the EE collects the source code location from the current state's syntactic information.

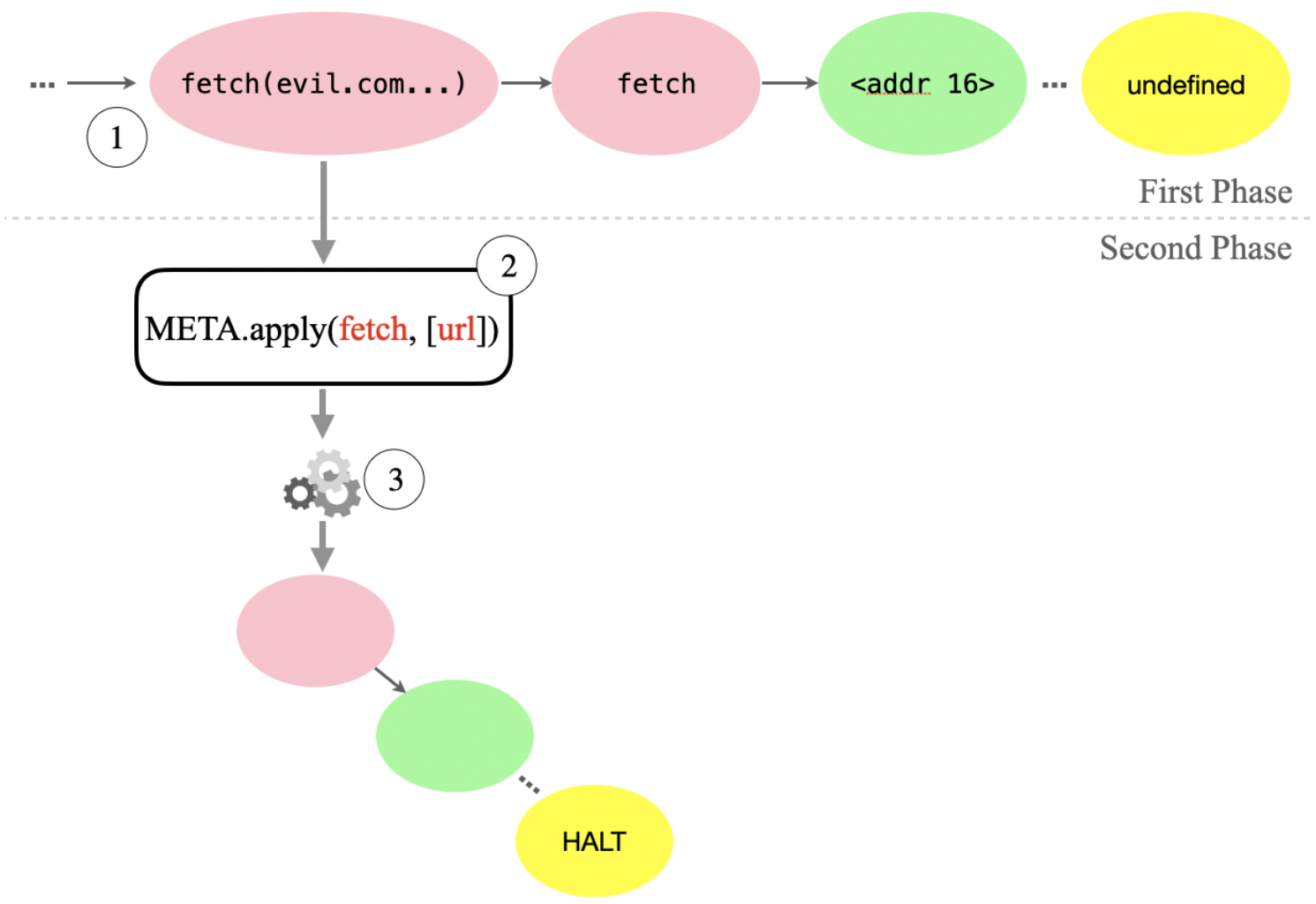

**Figure 3** Procedural view of the second phase abstract interpretation.

Central to our approach is the fact that the meta program is abstractly interpreted using a different configuration than the configuration used for the analysis of the base program. This is crucial for enabling suitable trade-offs between analysis speed and precision than when analysing the instrumented application in a single phase. Precision and speed can vary according to the development stage in which the static analysis component is used. For example, when used in an IDE, developers would like violations or bugs to be reported faster than when analysis is applied during a nightly build, when higher precision may be desired instead.

## 4 Phase 1: Static Analysis of Base Programs

We now elaborate on our two-phase abstract interpretation approach based on the calculus presented in prior work [34]. That work presented $\text{JS}_0$, a core functional language that models a subset of JavaScript, and a static analysis that models the execution of $\text{JS}_0$ programs as a flow graph from which information about control and value flow, and effects can be extracted. In this section, we present the syntax and semantics of $\text{JS}_0$ based on its original specification in [34]. We focus on the relevant features to understand the contributions of this work and include the details on the components and operations of the abstract machine in appendix D.

### 4.1 Syntax of $\text{JS}_0$

$\text{JS}_0$ is a core functional language that features objects as maps, higher-order functions, assignment, and prototype-based inheritance. The original syntax of $\text{JS}_0$ is depicted

$$
\begin{aligned}
e \in \mathsf{Exp} &::= s \mid f \mid v(s) \mid s_0.v(s_1) \mid \mathsf{new}\ v(s) \mid \mathsf{return}\ s \mid v\text{=}e \mid s.v \mid s.v\text{=}e \\
s \in \mathsf{Simple} &::= v \mid \mathsf{this} \\
f \in \mathsf{Fun} &::= \mathsf{function}\ (v)\{\mathsf{var}\ v_h;\ e\} \\
v \in \mathsf{Var} &= \text{a set of identifiers}
\end{aligned}
$$

**Figure 4** Input language $\text{JS}_0$.

in figure 4. Focusing on essential JavaScript features simplifies the presentation of our approach, although the features of $\text{JS}_0$ still are sufficient to show the applicability and generality of our approach. For validating our approach (section 6), we used an implementation of $\text{JS}_0$ that supports a substantially larger subset of JavaScript features such as conditionals, variable declarations, loops, exceptions, and type coercion. This implementation also defines parts of the standard built-in Javascript objects and functions.

In $\text{JS}_0$, all elements have a unique label to distinguish between different occurrences of the same syntactic expression (for example, to differentiate between the two references to pass in line 7 in listing 1).

### 4.2 Semantics of $\text{JS}_0$

We define the small-step semantics of $\text{JS}_0$ as an abstract machine [18] that transitions between states. This machine, based on the CESIK$^{\star}\Xi$ abstract machine introduced in Johnson and Van Horn [19], operates on abstract values but can be configured to express concrete semantics.

**State-space Components** The space-state of the abstract machine semantics is shown in figure 5. We now describe its main components. A machine state is either an evaluation state (**ev**) or a continuation state (**ko**). In an evaluation state, the machine evaluates an expression $e$ in an environment $\rho$. In a continuation state (**ko**), the machine is ready to continue evaluation with a value $text$ it has just computed. An environment $\rho$ maps variables to addresses. A store ($\sigma$) maps addresses ($a$) to values. In our formalism, we only consider two types of values: pointer values that correspond to addresses (i.e., pointers to an object), and primitive values **undef**, **true**, and **false**. Other primitive value types, such as numbers, strings, etc., can be added by extending the set *Prim*. $\mathscr{P}(X)$ denotes the *power domain* of set $X$. An object ($\omega$) is represented as a map from properties to values. Two internal object properties “call” and “proto”, distinct from regular properties, exist to implement function objects and object prototypes, respectively.

The stack is modeled as a combination of an intraprocedural continuation ($\iota$) and either an interprocedural continuation ($\kappa$) or the empty continuation ($\epsilon$). Interprocedural continuations play the role of *execution contexts* that are generated at call sites. An intraprocedural continuation also serves as a stack address pointing to underlying

$$
\begin{array}{rll}
\varsigma \in State & ::= \mathbf{ev}(e, \rho, \sigma, \iota, \kappa, \Xi) & \text{[eval state]} \\
& \mid\ \mathbf{ko}(text, \sigma, \iota, \kappa, \Xi) & \text{[kont state]} \\
\rho \in Env & = \mathsf{Var} \rightharpoonup Addr & \text{[environment]} \\
\sigma \in Store & = Addr \rightharpoonup (D + Obj) & \text{[store]} \\
text \in D & = \mathscr{P}(Addr + Prim) & \text{[value]} \\
\delta \in Prim & = \{\mathbf{undef}, \mathbf{true}, \mathbf{false}\} & \text{[primitive value]} \\
\omega \in Obj & = (\mathsf{Var} \rightharpoonup D) \times (\text{“proto”} \mapsto D) & \\
& \quad \times (\text{“call”} \mapsto \mathscr{P}(Callable)) & \text{[object]} \\
c \in Callable & ::= (f, \rho) & \text{[callable]} \\
\iota \in LKont & = Frame^{*} & \text{[frame]} \\
\phi \in Frame & ::= \mathbf{as}(v, \rho) & \text{[assignment frame]} \\
& \mid\ \mathbf{st}(s, v, \rho) & \text{[property store frame]} \\
\kappa \in Kont & ::= (e, c, text_{\text{arg}}, a_{\text{this}}, \sigma) & \text{[meta-continuation]} \\
\Xi \in KStore & = Kont \rightharpoonup \mathscr{P}(LKont \times Kont) & \text{[stack store]} \\
a \in Addr & \text{is a set of addresses} & \text{[address]}
\end{array}
$$

**Figure 5** State-space of the abstract machine semantics.

stacks in a stack store ($\Xi$). The empty continuation corresponds to the root context, which is created at the start of the program evaluation.

**Transition Relation** The semantics of the different syntactic elements of $\text{JS}_0$ are implemented as transition rules from evaluation states (**ev**). For example, the semantics of a method call, implemented by the E-METHOD-CALL rule, is applied whenever the machine reaches a state where $e$ is a method call expression. Transition rules may also use auxiliary relations or evaluation functions. For example, the function *evalSimple* is used to evaluate different types of simple expressions in $\text{JS}_0$, like literals, references and the *this* expression. The full specification of $\text{JS}_0$ rules, auxiliary functions and relations can be found in appendix D.

**Concrete and Abstract Evaluation** A program can be evaluated by calling the function eval with three arguments: eval($e$, $\alpha$, *alloc*). $e$ is the expression or program to evaluate. This expression is injected into an initial evaluation state, from which all other reachable states will be computed by the abstract machine. Function $\alpha$ converts concrete values into abstract values. Function *alloc* is used for generating addresses for allocation into the value store ($\sigma$). The result of evaluating $e$ is a flow graph in which nodes are reachable states, and edges are transitions between states. The flow graph, therefore, represents the steps taken by the machine during the abstract interpretation.

Functions $\alpha$ and *alloc* can be used to control the precision of evaluation and enable the machine to express concrete and abstract semantics.

For *concrete semantics*, function $\alpha$ is the identity function, so that the machine operates on concrete values. Additionally, the store allocator function $\sigma$ must always return fresh (i.e., unused) addresses. This configuration yields an abstract machine that computes a concrete interpretation of the given program. Analysing a program using concrete semantics is equivalent to the execution of the same program in a standard interpreter.

Because concrete semantics can yield large or infinite executions (and, therefore, large or infinite flow graphs), *abstract semantics* are used when performing static analysis to guarantee finite and tractable flow graphs. In abstract semantics, abstraction function $\alpha$ maps concrete values onto elements of a (bounded) lattice, which represents the (finite) abstract domain of those values. For example, function $\alpha$ can map concrete values to their type. The store allocator function $\sigma$ additionally must choose addresses from a finite set and may therefore return addresses that are already used in the store.

To illustrate parameterization for concrete and abstract semantics, consider again the base program in listing 4 as an example. Analysing this program using concrete semantics with variable `flags` equal to `true`, results in the graph shown in figure 1. The flow graph only includes the states reached by the machine during the analysis of `if` statement test expression and the consequent branch (which is the branch taken by the machine after the test expression).

As the figure shows, transition rule E-SIMPLE (defined in figure 9 in appendix D.2) is applied to state 7, meaning that the program is about to perform an evaluation of a simple expression (i.e., identifier operation in this case). The transitions relations of our abstract machine (described in appendix D.2) are expressed in sufficiently small steps, allowing us to obtain a very detailed graph in which program operations can be easily identified. Figure 2 is the flow graph resulting from the analysis of listing 4 using abstract semantics. As figures 1 and 2 make clear, different machine parameterizations influence the number of reachable states and the transitions between these states, but not the different kind of program operations and the manner in which they can be detected in the resulting flow graph.

## 5 Phase 2: Static Analysis of Meta Operations

The second phase of our *two-phase* (2PH) approach explores the flow graph resulting from the first phase to perform the *target analysis* embedded in the meta program (i.e., policy enforcement code for verifying AC and IFC policies). Exploring the flow graph for evaluating the appropriate meta program operations is the Execution Explorer (EE) responsibility, first described in section 3.2. The same abstract machine evaluator used for generating the flow graph in the first phase is used for abstractly executing the meta code (which is also JavaScript). The semantics of $\text{JS}_0$ therefore not only form an *operational* foundation for a static analysis of the base program but also for a *result-oriented* abstract interpretation of the meta code in this second phase to determine the outcome of the analysis (appendix E).

Although the same abstract machine is used for performing abstract interpretation, the key idea of splitting the abstract interpretation into two phases is to enable the use of a different parameterization for each individual phase. In terms of code size and complexity, the base program is usually significantly larger, and therefore typically is analyzed with lower precision than the meta program. The meta program containing the analysis code, on the other hand, is usually significantly smaller and less complex than the base program and is also under tighter control of the analysis developer. To remain faithful to the intended analysis semantics, the meta code therefore, typically is evaluated with significantly higher precision than the base code. In fact, in our experiments we configured the second phase to be as precise as possible. More concretely, this configuration is as close as possible to concrete semantics (e.g., with full precision), only losing precision due to control and value flow imprecisions introduced in the first phase.

In the rest of this section, we describe how our analysis intercepts base program operations and invokes meta program operations (i.e., performs the target analysis) from a given flow graph. We also discuss the treatment of a meta store for enabling stateful analysis of stateful policies.

### 5.1 Intercepting Base Program Operations and Invoking Traps

The second phase involves exploring the flow graph that resulted from the first phase. The Execution Explorer (EE), which is the static counterpart of the Execution Monitor (EM) for RASP, visits every state in the flow graph. For every state, the EE checks whether it represents a program operation that is trapped. In our running example, this may be calls to functions `check`, `fetch`, etc. in listing 1. Other examples include reading and writing object properties.

Intercepting base program operations is relatively straightforward when looking into the transition relation for $\text{JS}_0$ (shown in appendix D.2). Considering the interception of function calls as an example, we can observe that states in which state transition rules E-FUN-CALL, E-METHOD-CALL, and E-CTR-CALL apply, are states in which a function is about to be called. For example, state 17 in figure 2 is such a state. In our semantics, we define a relation *handle* that the EE applies to every state when exploring a base program's flow graph. This relation is formalized in appendix E.2. Relation *handle* takes a state and a meta store (we explain the latter below) and checks whether there is a trap method that corresponds to the evaluation step represented by that state; if so, it executes the trap method. The result of executing a trap method is a tuple consisting of the return value (`META.HALT` or `META.PROCEED` for our AC and IFC security analyses) and a meta store. We discuss obtaining and executing trap methods next.

#### 5.1.1 Obtaining Trap Methods

If a trapped operation is detected for a state, relation *handle* first obtains the reference to the `META` handler object that represents the access point to the analysis code. Recall that in our implementation, we make `META` a property of the global object, so obtaining this reference amounts to performing a straightforward property lookup on the global object within a state. Next, *handle* looks up the trap method that corresponds to the trapped operation by using the name of the trapped operation (e.g., apply, get, set, etc.). Looking up this method again amounts to a property lookup, this time on the META handler object that is reachable in each state. Continuing the previous example, if in state 17 in figure 2) a function call is intercepted, then *handle* will look up method META.apply in that state. The step of obtaining a trap method is formalized as the *trap* relation, given in appendix E.1. In particular, Trap-Callable specifies how to obtain the trap method META.apply.

#### 5.1.2 Executing Trap Methods

Finally, relation *handle* will abstractly execute the trap method, which corresponds to invoking the associated meta operation of the handler object. Remember that, even though the input base program application already *included* META as a library, the meta operations were never *called* during the first phase. It is only in the second phase that the abstract execution of the meta code (i.e., the enforcement of the security policy) is performed by obtaining and abstractly executing the appropriate trap methods. If the value resulting from executing a trap method subsumes the abstract value META.HALT, then this indicates that, according to the target analysis, a base program operation was intercepted that must halt the execution.

Since the abstract interpretation of trap methods is not an "analysis" but rather result-oriented abstract execution, trap methods are always executed with the highest possible precision that is bounded by the precision obtained during the abstract interpretation of the first phase. The reason for that is that the resulting precision of each abstract execution of each trap method is affected by the imprecision of the base program analysis. For example, analysing META.apply({{set:70}}, {Str}) in the second phase may introduce imprecision if the second argument ($\{Str\}$) is used.

The abstract execution of the same trap methods that are used in the dynamic analysis is what makes our approach *safe* and *efficient* because the same analysis specification (meta code) and semantics (abstract machine) are used for both the static and dynamic analysis.

### 5.2 Maintaining Analysis State

So far, we have ignored the issue of stateful analysis code, i.e. a meta program that maintains state of its own by performing side-effects on it. The AC policy enforcement code in listing 3 is an example of a stateful analysis because it performs side-effects to update a counter variable. Likewise, the IFC policy enforcement from listing 7 is stateful because it maintains a shadow stack.

In case the analysis is stateless, it suffices for the Execution Explorer (EE) to visit all flow graph states once in an unspecified order and handle these states as described in section 5.1. In case the analysis is stateful. However, meta state has to be maintained as well. The meta store is the component that is responsible for maintaining the meta state in our approach. As mentioned before, the relation *handle* takes a state and a meta store as input and returns a value and a meta store. The input meta store represents the meta state that a trap method has access to, and any modifications that a trap method makes to this meta state are captured in the returned meta store.

At the start of a trap method's execution, the meta store is merged into the base store contained in the state that represents the trapped operation. When the trap method returns, the meta store is obtained by collecting all heap information reachable from the META object. Function $\mathcal{R}$ in appendix E.2 formalizes the concept of reachability in a store.

Any meta state changes resulting from the execution of trap methods in a certain state $\varsigma$ must be propagated to subsequent executions of traps in states reachable from $\varsigma$. Because the flow graph may contain cycles, this means that handling states using *handle* now must be expressed as a fixed point computation over the flow graph. The fixed point is reached when handling each state adds no new information to either the return value or the meta store for that particular state in comparison to the previous call to *handle*.

In our semantics, the transition rule EE-Trans, detailed in appendix E.2, formally describes a transition from a reachable triple representing a state, a trap's return value, and a meta store, to triples for successor states in the flow graph. The transitive closure of this relation represents the fixed point that the EE computes.

## 6 Evaluation

In this section, we evaluate the applicability of our overall solution for deriving a SAST component from a RASP component (section 6.1) and empirically compare the precision and performance of the two-phase abstract interpretation approach compared to the single-phase approach (section 6.2).

**Setup** We use JIPDA [34] as a configurable abstract interpreter to perform static analysis of JavaScript programs and abstractly execute policy code. We employed JIPDA because it is possible to steer the abstract interpreter for implementing our 2-phase approach, and it implements a version of the syntax and semantics presented in section 4. Therefore the evaluation of a JavaScript program with JIPDA outputs a flow graph that approximates the behavior of that program for all its possible execution paths. JIPDA and the flow graphs it produces fulfill the assumptions discussed before. For RASP, we use a revised version of Guardia [41] for AC policies, which we load as a library before analysing a particular application.

Overall, the kind of JavaScript applications that our integrated RASP and SAST toolchain can analyse is tight to the underlying source code instrumenter employed in the RASP component. In our current prototype, this means JavaScript code mostly compatible with ECMAScript 5 as we currently do not support the following features: import/export, classes, generator functions, promises, and async/await functions.

### 6.1 Evaluation of Applicability

We evaluated the applicability of our approach by statically verifying a set of Access Control (AC) and Information Flow Control (IFC) policies from the literature. The goal is to demonstrate that our approach is general enough to dynamically enforce and statically verify the two most well-known types of security policies starting from only the specification of the policies and without requiring the reimplementation of the enforcement code for the static analysis.

#### 6.1.1 Access Control Policies

We verified a set of twelve AC policies from the literature summarized in table 2 (in appendix A). The policies were defined in a revised version of GUARDIA [41]. An example of how programmers would write a policy to *"prevent dynamic creation of iframe elements"* is shown in listing 8.

**Listing 8** Example policy written with GUARDIA (revised version) to prevent dynamic creation of iframe elements.

```
GG.onCall(iframe).with(GG.arg(GG.equals(0,'iframe'))).deny()
```

For each of the policies in table 2, we designed a test program that attempts to bypass the policy. Each test program was first executed with the RASP component. Then the derived SAST component using our two-phase approach was executed on the same set of programs to determine the number of policy violations it would detect. We found that the SAST component effectively detected all the intended policy violation for all tests.

#### 6.1.2 Information Flow Control Policies

table 3 shows the results of analyzing 14 programs without hidden implicit flows, out of 28 IFC test programs in Sayed et al. [39]. The derived SAST component was able to detect policy violations in all 14 test programs. We believe that support for implicit control flows would improve the security guarantees of our IFC monitor. However, it has been shown previously [43] that tracking explicit flows is sufficient to enforce *integrity* IFC policies, while for privacy-related IFC policies tracking hidden flows may be required. Therefore, from a software development point of view, the application developer can use our two-phase approach to know when sensitive components in an application may be affected by untrusted data (i.e., for integrity).

### 6.2 Evaluation of Performance and Precision

In this section, we validate our *two-phase* static analysis approach (2PH) by comparing it with the analysis of the instrumented version of the application (1PH) in terms of precision and analysis speed. The goal of this evaluation is two fold:

- Validate a key property of our approach: when the runtime enforcement of security policies is provided through metaprogamming in JavaScript, the static verification can be automatically obtained through a two-phase abstract interpretation approach, without the need to analyze the entire instrumented program.
- Confirm our hypothesis that, under the same analysis configuration, two-phase 2PH performs better than 1PH in terms of speed and precision.

**Methodology** We evaluate the performance and precision based on 8 small experimental client-side web applications, ranging from 46 to 110 LOC of HTML and JavaScript.

All applications were subject to the policy in listing 2. Despite the experimental applications being small compared to real-world applications, they do contain a substantial set of features from JavaScript and browsing environment (including DOM elements and events). Because the applications are not deliberately vulnerable, we randomly inserted calls to `fetch()` in their source code to perform the experiments. For each experimental application, we generated an instrumented version as a counterpart in which all function calls are instrumented.

For the analysis of uninstrumented applications using our 2PH, we use a tunable lattice that allows changing the precision of the abstraction function $\alpha$ (i.e., *high* and *low*) during the analysis. However, during an application analysis, the analysis program will lower the lattice's precision based on a threshold to produce values that are more abstract to ensure termination. Each uninstrumented application was analyzed twice using the 2PH approach. First, the application was analyzed using a high precision $\alpha$ for the first stage of our approach. For the second run, the application was analyzed with a low precision $\alpha$ for the first stage of our approach. Independent of the $\alpha$ function used during the first stage of the analysis, we always use a high precision $\alpha$ function for the second phase of our 2PH. The instrumented applications were also analyzed twice. For the first analysis, we used a high precision $\alpha$ for the lattice, while for the second, we used a low precision $\alpha$.

#### 6.2.1 Precision

Table 1 shows the results of the analysis of all experimental applications using the configurations explained before. Each row of the table is the result of analyzing a *program*, using an *approach* (i.e 1PH or 2PH). For the 1PH approach the instrumented version of the benchmark program was analyzed. As shown in the table, 2PH outperforms 1PH for all pairs of equivalent applications using the same lattice configuration. This experimentally validates the claim that a two-phase abstract interpretation for statically analyzing a base program (phase 1) and its instrumentation (phase 2) enables a better trade-off between precision and performance than a single static analysis of the instrumented code. This is because our 2PH approach runs with a high precision configuration in the second phase for the lattice. We did not observe any false negatives during our experiments because the experimental programs are simple. However, false negatives can occur in larger or more complex programs given the fact that our analysis is unsound. As mentioned before, practical static analyses of non-trivial JavaScript applications using dynamic features is always unsound to some degree, so we envision our 2PH approach to be used early during the development phase, or as part of a building pipeline to help catch vulnerabilities introduced by developers of an application, or the third-party code they include.

#### 6.2.2 Performance

To measure performance, we use the same setup as for measuring precision. Figure 6 shows the analysis speed in seconds for all applications using the four different combinations ($2PH_H, 2PH_L$, $1PH_H, 1PH_L$). The broken bars show the analyses that did not finish in a predefined time window (430 seconds).

**Table 1** Precision comparison between the single-phase approach (*1PH*) and our two-phase approach (*2PH*) for statically detecting AC policy violations. Column *Precision* indicates the analysis precision: *H* for high precision, *L* for low precision. Columns *TP*, *FP*, and *FN* denote the number of true positives, false positives, and false negatives, respectively, with respect to reported policy violations by each approach. "-" denotes the absence of a value due to analysis timeout.

| Program | Precision | 1PH | | | 2PH | | |
|---|---|---|---|---|---|---|---|
| | | TP | FP | FN | TP | FP | FN |
| sequential | H | 3 | 0 | 0 | 3 | 0 | 0 |
| | L | - | - | - | 3 | 0 | 0 |
| branches | H | 0 | 4 | 0 | 0 | 4 | 0 |
| | L | 0 | 7 | 0 | 0 | 4 | 0 |
| iterative | H | 0 | 3 | 0 | 0 | 0 | 0 |
| | L | 0 | 4 | 0 | 0 | 2 | 0 |
| safe | H | 0 | 4 | 0 | 0 | 0 | 0 |
| | L | 0 | 6 | 0 | 0 | 2 | 0 |
| recursive | H | - | - | - | 1 | 0 | 0 |
| | L | 1 | 3 | 0 | 1 | 0 | 0 |
| fib | H | - | - | - | 1 | 0 | 0 |
| | L | 1 | 3 | 0 | 1 | 0 | 0 |
| passStrength | H | - | - | - | 2 | 1 | 0 |
| | L | - | - | - | 2 | 3 | 0 |
| steal | H | 0 | 3 | 0 | 0 | 0 | 0 |
| | L | - | - | - | 0 | 1 | 0 |

As shown in the figure, the 2PH approach performs better than the 1PH for both lattice configurations. Performance is influenced by how the fixed point is computed, which in the second phase is affected by only the meta store. There is a similar performance for the configurations $1PH_H$ and $2PH_H$ for the *sequential, branches, iterative* and *safe* programs. Most likely, this is because the values remained under the configured threshold for the lattice during the analyses. From our experimental data, we also observe that high analysis precision often results in better performance than low analysis precision. Although the speed–precision trade-off for a static analysis depends on many factors and is generally unpredictable, lowering precision tends to slow down an analysis because this increases the number of spurious paths explored. However, more experimental data is needed to draw a general conclusion about this.

The performance can also be measured in function of the number of states generated during the analysis. Figure 7 shows the relation of states produced by the 1PH and 2PH approaches for our set of programs evaluated in both high and low precision lattice configurations. As the figure shows, our 2PH approach performs better than the 1PH for most of the programs in terms of states generated (with more states representing more explored paths). For the examples *iterative* (H and L), *recursive* (L) and *fib* (L) the 1PH performs slightly better than our 2PH, however, at the cost of precision. The broken bars means that the analysis generated more than 100000 states.

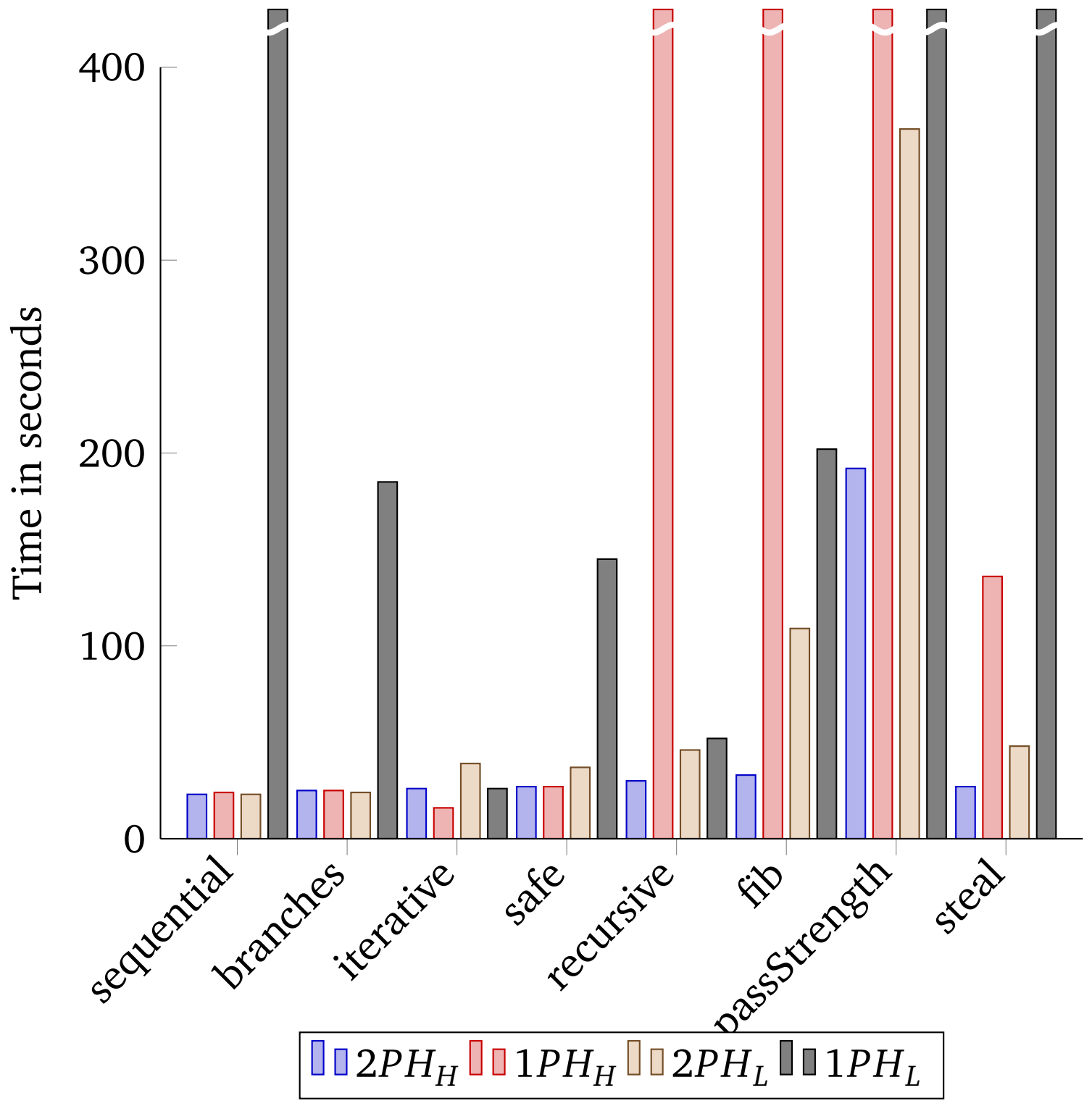

**Figure 6** Speed comparison between the *1PH* approach and our (*2PH*) approach for statically detecting AC policy violations with high precision ($H$) and low ($L$) precision. Each application is thus analysed using four different configurations ($2PH_H$, $2PH_L$, $1PH_H$, $1PH_L$).

## 7 Discussion

We now discuss some of the properties of our two-phase abstract interpretation approach for deriving SAST from RASP, as well as future work.

**Efficiently Deriving SAST** Our goal is to derive a static analysis from an already existing dynamic analysis mechanism with the least effort. Although our two-phase approach avoids the reimplementation of the meta program (i.e., policy enforcement code) when deriving the static analysis from the dynamic enforcement, the triggering mechanism used in the second analysis phase is not reused from the dynamic enforcement mechanism and therefore still has to be reimplemented. However, typically the policy code is much larger and more complex than the triggering code. For instance, the code for monitoring a specific application behavior, such as function calls and value creation, is similar in both structure in the runtime and the static implementation. Therefore, we believe that the advantages of performing static analysis in two phases as described in section 3.2 outweighs the downside of having to reimplement (only) the triggering mechanism.

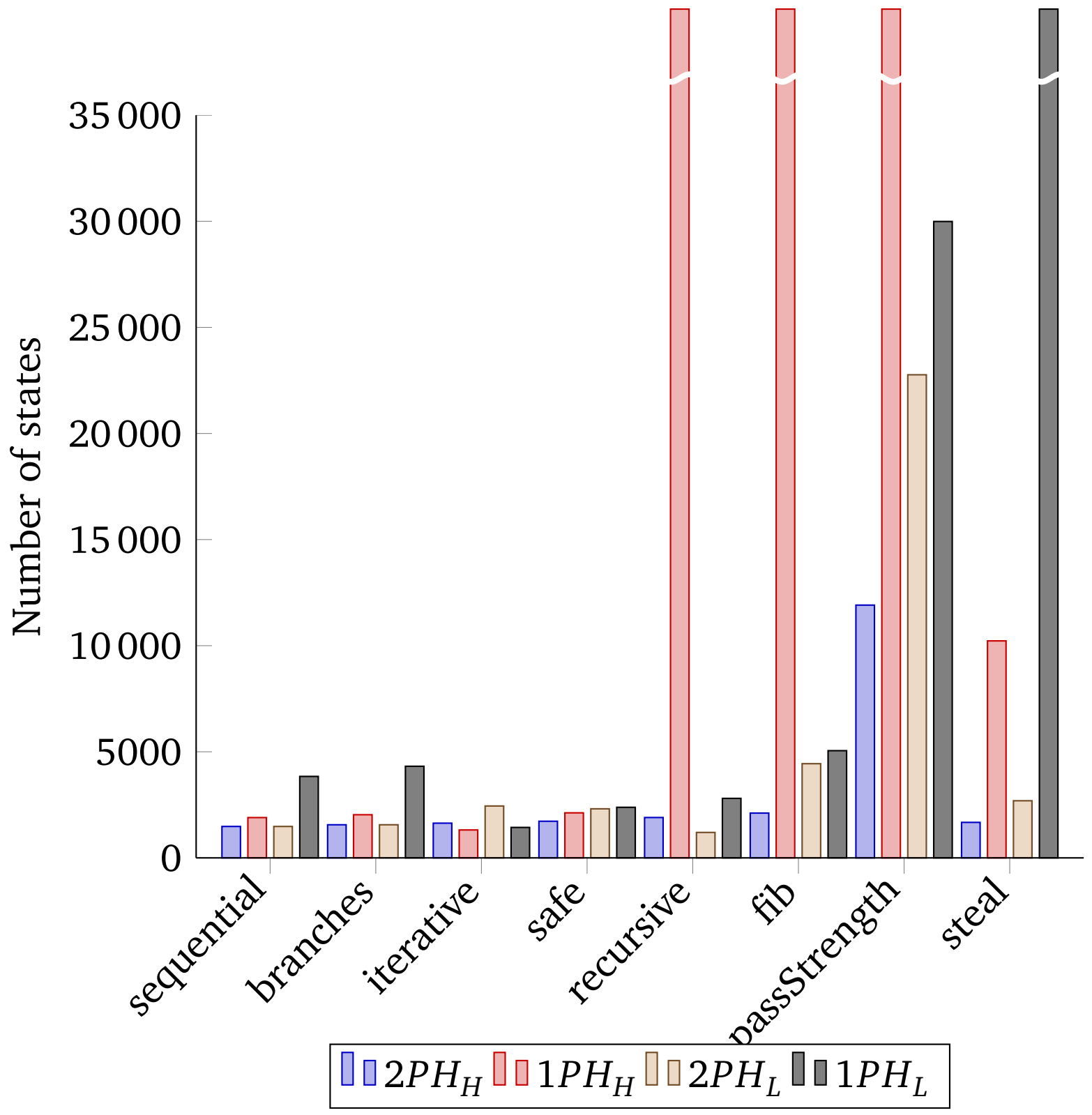

**Figure 7** Comparison between the number of states generated by 1PH approach and our 2PH during analysis using *low* (L) and *high* (H) precision lattice configurations. Each application is thus analysed using four different configurations ($2PH_H, 2PH_L$, $1PH_H, 1PH_L$).

Performing static analysis in two phases can also bring benefits in terms of speed and precision (as we showed in section 6). Since precision is one of the most important properties of any static analysis, it deserves some discussion. We would like to emphasize that our goal is not to improve the precision of our 2PH static analysis over other custom static analyses. Instead, the goal is to avoid writing the policies' semantics for the static analysis by reusing the policies' semantics already implemented by the runtime monitor. As explained in section 5, the precision of our analysis depends on the flow graph resulting from the first phase. As such, abstracted base program values in the first phase can introduce imprecision to our security analysis during the second phase.

*What* the meta code does is independent of the static analysis. In contrast, it does matter *how* the meta code is implemented. Therefore, it is a reasonable assumption that analysis developers should avoid features that the abstract interpreter cannot soundly or precisely handle (e.g., `eval`, the `Function` constructor, etc.).

**Implicit Flows** As mentioned in section 3.1, the IFC monitor used in this work cannot handle language features that cause hidden implicit flows. We would like to stress that this is not a limitation of our two-phase approach but a limitation of the employed IFC monitor. Actually, this is a challenge for dynamic IFC monitors in general. In previous work, we developed an IFC monitor, GIFC [40], which covers implicit flows. While GIFC is an example of a security policy library that instruments a program by trapping operations so they can be checked against an IFC policy, this cannot be used as a RASP component because its implementation relies on JavaScript proxies which are not currently supported in JIPDA.

Other work that has explored the precise handling of hidden implicit flows in an IFC monitor relies on a static analysis for control flow statements prior to the execution of the program [6, 12]. Combining such a technique into our approach requires adding yet another static analysis phase which is not trivial. For example, statements such as `continue`, `break`, and non-tail `return` create implicit flows when they are used conditionally in expressions that include tainted information. It would be interesting to explore a solution to support IFC policies that at runtime also require a static approximation of the program runtime behavior to correctly deal with hidden implicit flows.

**Applicability to Other Analyses beyond Security** We believe that the two-phase abstract interpretation approach presented in this paper is more broadly applicable than just security. As future work, we would like to explore its applicability to other dynamic analysis that can be implemented with runtime monitors based on metaprogramming akin to the RASP components used in this work. For example, Gong et al. [21] developed a dynamic linter (DLint) to detect violations of code quality rules dynamically. In their framework, a rule is specified as a predicate over program events (i.e. function calls, variable writes, etc.) to check common quality issues in JavaScript, like inheritance and type inconsistency problems, APIs misuse, etc. Another kind of dynamic analyses that could be implemented are profilers. For example, JITProf [20] is a profiling framework to dynamically identify code locations that prohibit JIT optimizations. In JITProf, the information of program events is used to maintain metadata associated with the source location of those events. The metadata is then used to pinpoint those locations that prohibit JIT (i.e., just-in-time) optimisations.

## 8 Related Work

To the best of our knowledge, this is the first work that derives a static analysis mechanism from a dynamic security policy enforcement mechanism. In this section, we discuss previous efforts to unify or combine static and dynamic analysis.

JNuke [3] integrates static and dynamic analysis through an architecture that enables the use of a generic algorithm in both contexts. The dynamic analysis instrument the virtual machine to record a trace of events of interest. On the other hand, static analysis is implemented as a graph-free analysis through abstract interpretation, which allows the analysis to be *close* to a dynamic execution. To achieve the integration, the framework executes the algorithm in an environment that abstracts the differences between the program states generated from static and dynamic environments on which the application is executed.

Bodden et al. [8, 10] develop a system to verify runtime monitors ahead of time. Their solution is tailored to the syntax and semantics of the language used to express the monitor. In contrast, our SAST component is agnostic and can therefore support different policy specification languages and their semantics (e.g., the AC and IFC policy specifications described in section 6) without additional effort.

The CLARA [9] framework enables the partial evaluation of finite-state runtime monitors ahead of time. CLARA's static analysis can detect possible violations that may occur at runtime. Its monitor only observes events triggered by program locations for which the static analysis failed to prove safety. In contrast to our approach, CLARA does not derive static analysis from the monitor. However, it applies a series of ad-hoc pluggable static analyses to iteratively prune unnecessary instrumentation points.

Tripp and Weisman [46] develop a hybrid analysis for performing a security assessment of client-side JavaScript code. In their approach, a web crawler retrieves and executes a web page to record useful runtime information. The resulting JavaScript code and recorded information are passed to a static taint analysis component for performing the actual security assessment. Hybrid analyses such as this one combine static and dynamic analysis in such a way that they rely on each other for their operation. In contrast, the dynamic enforcement in our approach can be used independently from any static analysis. Once a static analysis is derived, it can be applied without executing the dynamic enforcement.

Some related work that targets the development stage of the software development cycle complements static verification with dynamic information, e.g. [44, 47]. The number of false alarms produced by such an analysis could, in theory, be reduced since they have runtime information available during static analysis. However, their ability to improve the reporting of violations is constrained to the program coverage offered by the dynamic information and the actual runtime information that is recorded. Our approach covers the two extremes of the software development cycle (SAST and RASP), and as such, policies are verified ahead of time and are also enforced at runtime.

Ahrendt et al. [1, 2] developed a framework that unifies static and runtime verification of Java applications. The instantiation of their framework unifies two unrelated tools (KeY, a deductive verification system, and Larva, a logical automaton for runtime verification) through a unified language (*ppDATEs*) specially designed for the specification of static and runtime properties. The final instrumented application benefits from the proofs obtained by the static verification, enabling the monitor to make assumptions that prevent it from performing too many checks at runtime. In contrast, due to the unsound static analysis, our RASP component must always perform all checks.

Dufour et al. [16], describe a framework for blended analysis. The overall idea behind the framework is to apply an interprocedural static analysis on a calling structure obtained through dynamic analysis, therefore, capturing properties of a single execution. Wei and Ryder [47], use this framework to detect vulnerabilities on JavaScript applications through taint analysis.

Chugh et al. [14], develop a staged IFC analysis. In their approach, an offline analysis is performed in the application source code to compute a residual policy that is then enforced at runtime. The goal of the residual policy is to enforce the information flow policy in places where the application executes code dynamically generated (e.g. eval).

## 9 Conclusion

We presented an approach for performing both Static Application Security Testing (SAST) and Runtime Application Security Protection (RASP) using a single security policy library in a safe and efficient manner. Our approach starts from a set of declarative security policies, from which a runtime enforcement mechanism is generated by instrumenting the target application for trapping security-relevant program operations and forwarding them to policy enforcement code. Next, a static analysis mechanism is derived from the runtime enforcement mechanism without reimplementing any of the policy enforcement code. Our approach reduces the effort of combining SAST and RASP, offering strong guarantees that the policy semantics are identical between their static verification and their runtime enforcement. We demonstrated the applicability of our approach by detecting and enforcing 12 access control and 14 information flow control policies originating from related work.

## A Access Control Policies

The table 2 list a set of access control policies appearing in the related of language-based security mechanisms for client-side web applications.

**Table 2** Description of the 12 AC policies from related work for which we derived a static verification from their dynamic enforcement.

| Policy | Description |
|---|---|
| 1 | limit the number of popup windows [24] |
| 2 | disallow popup windows without location and status bar [24] |
| 3 | prevent abuse of resources like modal dialogues [24] |
| 4 | disallow dynamic iframe creation [37] |
| 5 | disable page redirects after document.cookie read [24] |
| 6 | only redirect to whitelisted URLs [37] |
| 7 | restrict XMLHttpRequest (XHR) to secure connections and whitelisted URLs [31] |
| 8 | disallow setting of src property on dynamic images [37] |
| 9 | disallow open and send methods of XHR object [37] |
| 10 | postMessage can only send to the whitelisted URLs [31] |
| 11 | disallow string arguments to setInterval [31] |
| 12 | disable geoposition API [41] |

## B Information Flow Control Policies

Table 3 shows the IFC test cases originated from the related work used during the evaluation of our 2PH approach.

## C Comparison of Number of States Generated by 1PH and Our 2PH Approaches

Table 4 relates the number of explored states of the 1PH and our 2PH under low (L) and high (H) lattice configurations.

## D Phase 1: Semantics of $JS_0$

Before delving into the details on the transition rules, auxilary functions and relations, we introduce the notation and conventions used.

**Notation and Conventions** We use $\uplus$ to denote *disjoint union*: if $X = Y \uplus Z$, then $Y = X \setminus Z$. The notation $X = x : X'$ deconstructs a *sequence* $X$ into its first element $x$ and the rest $X'$. We write $\langle\rangle$ for the empty sequence. The *power domain* of set $X$ is denoted as $\mathscr{P}(X)$. The *empty function* is denoted as $[]$, and for all inputs returns the bottom element $\bot$ of its range. The notation $f[x \mapsto y]$ denotes *function extension* and yields a function $f'$ such that:

$$f'(z) = \begin{cases} y & \text{if } z = x, \\ f(z) & \text{else.} \end{cases}$$

**Table 3** Result of applying the SAST component derived from our RASP IFC monitor on 13 test cases without hidden implicit flows from of Sayed et al. [39]. Each test case contains an IFC policy violation, and a checkmark in column *Violation detected* signifies that the static verification correctly detected this. Column *Features* lists the set of notable features are present in each test program: *if*—if statement, *lp*—for or while statement, *ret*—(conditional) return statement, *thr*—throw statement, *this*—this expression, *new*—new expression, *arr*—arrays, *oprop*—access or modification of object property, *oproto*—access or modification of prototype property.

| Test case [39] | Features | Violation detected |
|---|---|---|
| 1 | | ✓ |
| 2 | if | ✓ |
| 3 | lp | ✓ |
| 4 | lp | ✓ |
| 5 | lp | ✓ |
| 6 | lp, if, arr, oprop | ✓ |
| 7 | lp, if, cb, oprop | ✓ |
| 14 | oprop, this | ✓ |
| 15 | oprop, this, new | ✓ |
| 17 | oprop, this, new | ✓ |
| 19 | ret, oprop, this, new | ✓ |
| 22 | oprop | ✓ |
| 24 | ret, oprop, oproto, this, new | ✓ |
| 28 | lp, oprop | ✓ |

We write the *function restriction* (or narrowing) of a function $f$ to domain $X$ as $f|_X$, such that $(f|_X)(x) = f(x)$ if $x \in X$ and $(f|_X)(x) = \bot$ else. *Function joining* happens in a pointwise fashion. If $\sqcup$ is the join operator for the range of the function, then $[x \mapsto y_1] \sqcup [x \mapsto y_2] = [x \mapsto y_1 \sqcup y_2]$. In particular, $\bigsqcup\{[x_0 \mapsto y_0], \ldots, [x_n \mapsto y_n]\} = [x_0 \mapsto y_0] \sqcup \ldots \sqcup [x_n \mapsto y_n]$.

### D.1 Auxiliary Evaluation Functions and Relations

The evaluation rules for simple expressions are shown in figure 8. Function *evalSimple* : Simple $\times$ *Env* $\times$ *Store* $\times$ *Kont* $\mapsto$ $D$ evaluates three types of simple expressions in $\text{JS}_0$: literals, references to either a global or non-global variable, and this expression.

**Table 4** Comparison the between number of states generated by (*1PH*) and our (*2PH*) during the analysis of experimental applications using low (*L*) and high (*H*) precision lattice configurations.

| Program | Precision | States generated | |
|---|---|---|---|
| | | 1PH | 2PH |
| sequential | H | 1899 | **1481** |
| | L | 3839 | **1481** |
| branches | H | 2033 | **1560** |
| | L | 4317 | **1560** |
| iterative | H | **1316** | 1636 |
| | L | **1435** | 2445 |
| safe | H | 2122 | **1726** |
| | L | 2383 | **2312** |
| recursive | H | - | **1902** |
| | L | **1199** | 2842 |
| fib | H | - | **2111** |
| | L | **4439** | 5052 |
| passStrength | H | - | **11912** |
| | L | 29991 | **22767** |
| steal | H | 10266 | **1673** |
| | L | - | **2692** |

E-LIT
$$\frac{}{\mathit{evalSimple}(\delta,\rho,\sigma,\kappa)=\alpha(\delta)}$$

E-VAR
$$\frac{v\in \mathrm{Dom}(\rho) \qquad a=\rho(v)}{\mathit{evalSimple}(v,\rho,\sigma,\kappa)=\sigma(a)}$$

E-GLOBAL
$$\frac{v\notin \mathrm{Dom}(\rho) \qquad \omega_0=\sigma(a_0)}{\mathit{evalSimple}(v,\rho,\sigma,\kappa)=\omega_0(v)}$$

E-THIS
$$\frac{}{\mathit{evalSimple}([\![\mathsf{this}]\!],\rho,\sigma,(e,c,text_{\text{arg}},a_{\text{this}},\sigma))=\{a_{\text{this}}\}}$$

**Figure 8** Evaluation rules for simple expressions.

Relation *lookupProp* looks up a property by traversing the prototype chain of an object. If the property is not found in the chain, undefined is returned.

$$\mathit{lookupProp}(v,a,\sigma) = \begin{cases} \omega(v) & \text{if } v\in \mathrm{Dom}(\omega) \\ \{\mathbf{undef}\} & \text{if } \omega(\text{“proto”})=\varnothing \\ \mathit{lookupProp}(v,a',\sigma) & \text{else} \end{cases}$$

where $\omega=\sigma(a)$

$a'\in\omega(\text{“proto”})$

Function *evalCall* applies a callable (or closure) $(f, \rho)$ to an argument $text_{\text{arg}}$ in a certain program state. We assume a single parameter and a single local variable declaration. Therefore, *evalCall* extends the function's static environment $\rho$ and the store $\sigma$ by binding parameter to its argument value and the variable is hoisted to the beginning of the function scope and bound to undefined.

The continuation $\kappa$ is the execution context of the caller and parameter $\kappa'$ is the execution context for the call itself. The stack store $\Xi$ is extended by allocating the caller stack $(\iota, \kappa)$ at stack address $\kappa'$. The function's body evaluation occurs in the static environment and store extended with the binding of the argument ($\rho'$ and $\sigma'$), with an empty local stack $\langle\rangle$, and the execution context for the call $\kappa'$.

E-CALL

$$\frac{\begin{array}{c} f = [\![\text{function } (v)\{\text{var } v_h;\, e\}]\!] \\ \rho' = \rho[v \mapsto a] \qquad \sigma' = \sigma \sqcup [a \mapsto text_{\text{arg}}, a_h \mapsto \{\mathbf{undef}\}] \\ a = alloc(v, \rho, \sigma, \iota, \kappa) \qquad a_h = alloc(v_h, \rho, \sigma, \iota, \kappa) \qquad \Xi' = \Xi \sqcup [\kappa' \mapsto \{(\iota, \kappa)\}] \end{array}}{evalCall((f, \rho), text_{\text{arg}}, \sigma, \iota, \kappa, \Xi, \kappa') = \mathbf{ev}(e, \rho', \sigma', \langle\rangle, \kappa', \Xi')}$$

### D.2 Transition Relation

We define the transition relation $\mapsto$ of our abstract machine using the functions defined in the previous sections.

$$(\mapsto) \sqsubseteq State \times State$$

Rules for transitions from evaluation states (**ev**) correspond with the different syntactic cases, while rules for transitions from continuation states (**ko**) correspond with the different kinds of continuations. Figure 9 and figure 10 show the transitions rules.

### D.3 Program Evaluation

Function $\mathscr{I} : \text{Exp} \to State$ injects an expression into the state-space. It returns an initial evaluation state with empty environment, initial store, empty local continuation, and the root context ($\kappa_0$) as interprocedural continuation.

$$\frac{\kappa_0 = (e, \bot, \bot, a_0, \sigma_0) \qquad \sigma_0 = [a_0 \mapsto [\,]]}{\mathscr{I}(e) = \mathbf{ev}(e, [\,], \sigma_0, \kappa_0, \epsilon, [\,])}$$

The initial store $\sigma_0$ contains the global object at address $a_0$, which we assume to be available throughout the semantics.

If $\varsigma_0$ is the initial state for program $e$, then the evaluation of this program corresponds with computing the transitive closure of $\mapsto$ starting from $\varsigma_0$.

$$\frac{\kappa_0 = (e, \bot, \bot, a_0, \sigma_0) \qquad \varsigma_0 = \mathscr{I}(e) \qquad \varsigma_0 \mapsto^* \mathbf{ko}(text, _, \langle\rangle, \kappa_0, _)}{text \in \mathscr{E}(e)}$$

Our static analysis requires a finite model (i.e flow graph) for every possible (finite) program. This can be guaranteed by parameterizing the abstract machine with an

E-SIMPLE
$$\frac{text = evalSimple(s, \rho, \sigma, \kappa)}{\mathbf{ev}(s, \rho, \sigma, \iota, \kappa, \Xi) \mapsto \mathbf{ko}(text, \sigma, \iota, \kappa, \Xi)}$$

E-FUN-CALL
$$\frac{\begin{array}{c} text_f = evalSimple(v, \rho, \sigma, \kappa) \qquad text_{\text{arg}} = evalSimple(s, \rho, \sigma, \kappa) \\ a_f \in text_f \qquad \omega_f = \sigma(a_f) \qquad c \in \omega_f(\text{“call”}) \qquad \kappa' = (e, c, text_{\text{arg}}, a_0, \sigma) \end{array}}{\mathbf{ev}([\![ \underbrace{v(s)}_{e} ]\!], \rho, \sigma, \iota, \kappa, \Xi) \mapsto evalCall(c, text_{\text{arg}}, \sigma, \iota, \kappa, \Xi, \kappa')}$$

E-ASSIGN
$$\frac{\phi = \mathbf{as}(v, \rho)}{\mathbf{ev}([\![v\text{=}e]\!], \rho, \sigma, \iota, \kappa, \Xi) \mapsto \mathbf{ev}(e, \rho, \sigma, \phi : \iota, \kappa, \Xi)}$$

E-LOAD
$$\frac{text_r = evalSimple(s, \rho, \sigma, \kappa) \qquad a \in text_r \qquad text \in lookupProp(v, a, \sigma)}{\mathbf{ev}([\![s.v]\!], \rho, \sigma, \iota, \kappa, \Xi) \mapsto \mathbf{ko}(text, \sigma, \iota, \kappa, \Xi)}$$

K-ASSIGN-VAR
$$\frac{v \in \mathrm{Dom}(\rho) \qquad a = \rho(v) \qquad \sigma' = \sigma \sqcup [a \mapsto text]}{\mathbf{ko}(text, \sigma, \mathbf{as}(v, \rho) : \iota, \kappa, \Xi) \mapsto \mathbf{ko}(text, \sigma', \iota, \kappa, \Xi)}$$

K-ASSIGN-GLOBAL
$$\frac{v \notin \mathrm{Dom}(\rho) \omega = \sigma(a_0)[v \mapsto text] \qquad \sigma' = \sigma \sqcup [a_0 \mapsto \omega]}{\mathbf{ko}(text, \sigma, \mathbf{as}(v, \rho) : \iota, \kappa, \Xi) \mapsto \mathbf{ko}(text, \sigma', \iota, \kappa, \Xi)}$$

K-STORE
$$\frac{text_r = evalSimple(s, \rho, \sigma, \kappa) \qquad a \in text_r \qquad \omega = \sigma(a)[v \mapsto text] \qquad \sigma' = \sigma \sqcup [a \mapsto \omega]}{\mathbf{ko}(text, \sigma, \mathbf{st}(s, v, \rho) : \iota, \kappa, \Xi) \mapsto \mathbf{ko}(text, \sigma', \iota, \kappa, \Xi)}$$

K-CTR-RETURN
$$\frac{(\iota', \kappa') \in \Xi(\kappa) \qquad text' = \{a_{\text{this}}\} \qquad ([\![\texttt{new}\ v(s)]\!], _, _, a_{\text{this}}, _) = \kappa}{\mathbf{ko}(text, \sigma, \langle\rangle, \kappa, \Xi) \mapsto \mathbf{ko}(text', \sigma, \iota', \kappa', \Xi)}$$

K-FUN-RETURN
$$\frac{(\iota', \kappa') \in \Xi(\kappa)}{\mathbf{ko}(text, \sigma, \langle\rangle, \kappa, \Xi) \mapsto \mathbf{ko}(text, \sigma, \iota', \kappa', \Xi)}$$

**Figure 9** Transition rules of the abstract machine I.

address allocator that draws addresses from a finite set *Addr*. When both Var and *Addr* are finite sets, then the entire state-space is finite as well and $\mapsto$, which is monotonic, has a least fixed point.

E-FUN

$$\frac{\begin{array}{c} a = allocFun(f, \rho, \sigma, \iota, \kappa) \\ a' = allocProto(f, \rho, \sigma, \iota, \kappa) \qquad \sigma' = \sigma \sqcup [a \mapsto \{\omega_f\}, a' \mapsto \{\omega_{\text{proto}}\}] \\ \omega_f = [\text{“call”} \mapsto \{(f, \rho)\}, \text{“proto”} \mapsto \emptyset, \text{prototype} \mapsto \{a'\}] \qquad \omega_{\text{proto}} = [\text{“proto”} \mapsto \emptyset] \end{array}}{\mathbf{ev}([\![\underbrace{\text{function } (v)\{\text{var } v_h;\, e\}}_{f}]\!], \rho, \sigma, \iota, \kappa, \Xi) \mapsto \mathbf{ko}(\{a\}, \sigma', \iota, \kappa, \Xi)}$$

E-METHOD-CALL

$$\frac{\begin{array}{c} text_{\text{this}} = evalSimple(s_0, \rho, \sigma, \kappa) \qquad text_{\text{arg}} = evalSimple(s_1, \rho, \sigma, \kappa) \\ a_{\text{this}} \in text_{\text{this}} \qquad text_f \in lookupProp(v, a_{\text{this}}, \sigma) \\ a_f \in text_f \qquad \omega_f = \sigma(a_f) \qquad c \in \omega_f(\text{“call”}) \qquad \kappa' = (e, c, text_{\text{arg}}, a_{\text{this}}, \sigma) \end{array}}{\mathbf{ev}([\![\underbrace{s_0.v(s_1)}_{e}]\!], \rho, \sigma, \iota, \kappa, \Xi) \mapsto evalCall(c, text_{\text{arg}}, \sigma, \iota, \kappa, \Xi, \kappa')}$$

E-CTR-CALL

$$\frac{\begin{array}{c} text_f = evalSimple(v, \rho, \sigma, \kappa) \\ text_{\text{arg}} = evalSimple(s, \rho, \sigma, \kappa) \qquad a_f \in text_f \qquad \omega_f = \sigma(a_f) \\ c \in \omega_f(\text{“call”}) \qquad a_{\text{this}} = allocCtr(e, \rho, \sigma, \iota, \kappa) \qquad \omega = [\text{“proto”} \mapsto \omega_f(\text{prototype})] \\ \sigma' = \sigma \sqcup [a_{\text{this}} \mapsto \{\omega\}] \qquad \kappa' = (e, c, text_{\text{arg}}, a_{\text{this}}, \sigma) \end{array}}{\mathbf{ev}([\![\underbrace{\text{new } v(s)}_{e}]\!], \rho, \sigma, \iota, \kappa, \Xi) \mapsto evalCall(c, text_{\text{arg}}, \sigma', \iota, \kappa, \Xi, \kappa')}$$

E-STORE

$$\frac{\phi = \mathbf{st}(s, v, \rho)}{\mathbf{ev}([\![s.v\text{=}e]\!], \rho, \sigma, \iota, \kappa, \Xi) \mapsto \mathbf{ev}(e, \rho, \sigma, \phi : \iota, \kappa, \Xi)}$$

E-RETURN

$$\frac{text = evalSimple(s, \rho, \sigma, \kappa)}{\mathbf{ev}([\![\text{return } s]\!], \rho, \sigma, \iota, \kappa, \Xi) \mapsto \mathbf{ko}(text, \rho, \sigma, \langle\rangle, \kappa, \Xi)}$$

**Figure 10** Transition rules of the abstract machine 2.

## E Phase 2: A Posteriori Abstract Interpretation of Meta Operations

During the second phase, the appropriate meta program operations must be triggered by exploring the output of the abstract interpretation from the first phase. This is the responsibility of the Execution Explorer (EE) (cf. section 3.2). The semantics of $\text{JS}_0$ therefore not only form an *operational* foundation for a static analysis of the base program, but also for a *result-oriented* abstract interpretation of the meta code in this second phase. Both objectives are fulfilled by representing the semantics of $\text{JS}_0$ as an abstract machine that models evaluation in small steps.

Intercepting operations is relatively straightforward when looking at the transition relation for $\text{JS}_0$ (cf. appendix D.2). Again, taking the interception of function calls as an example, we can observe that rules E-FUN-CALL, E-METHOD-CALL, and E-CTR-CALL are states in which a function is about to be called.

Upon detection of an operation that must be intercepted, the EE has to invoke the associated meta operation on the handler object. However, instead of duplicating the behavior of the run time meta operation by extending the abstract machine of the static analysis, the EE relies on the meta operations defined on the run time handler (i.e META). At this point, the EE will initiate an abstract interpretation invoking the appropriate trap on the handler. This is possible because the base program already included the code of meta program, and the base and meta language are the necessarily the same (because the EM is based in source code instrumentation). This forms the crux of our approach.

### E.1 Obtaining the Callable Object

Suppose $\varsigma$ is a state in which an operation $v_m$ must be intercepted. The EE first has to obtain the handler object META from this state, which, as before, we assume is a property of the global object. Rule TRAP-CALLABLE obtains a reference $text_M$ for the handler object by looking up the [[META]] property on the global object with address $a_0$. The trap method is looked up as a property with name $v_m$ on the handler, resulting in a reference $text_m$ to a function object. Finally, the value of the "call" special property is returned.

TRAP-CALLABLE

$$\frac{\begin{array}{c} text_M \in lookupProp([[\mathsf{META}]], a_0, \sigma_\varsigma) \\ a_M \in text_M \qquad text_m \in lookupProp(v_m, a_M, \sigma_\varsigma) \\ a_m \in text_m \qquad \omega_m = \sigma_\varsigma(a_m) \qquad c_m \in \omega_m(\text{"call"}) \end{array}}{c_m \in trap(\varsigma, v_m)}$$

### E.2 Intercepting Base Program Operations and Invoking Traps

We define a relation *handle* that the EE uses when exploring the resulting base program's flow graph. Relation *handle* takes a state and a meta store, and invokes the required trap if required. It returns the result of a trap invocation and a resulting meta store. The meta store is required for maintaining meta state, and is explained below. Our explanation here focuses on interception of operations and trap invocation, and as an example we illustrate the rule for intercepting method calls and invoking the corresponding apply trap.

Suppose $\varsigma$ is a state that, upon transition, results in a method call, i.e., transition rule E-METHOD-CALL in figure 10 applies to $\varsigma$. From the specification of the EM the apply trap has to be invoked. The EE encode this behavior by means of the rule Handle-Method-Call for relation *handle*.

$$\textsc{Handle-Method-Call}\quad \frac{\begin{array}{c}\mathbf{ev}([\![s_0.v(s_1)]\!],\ldots)=\varsigma \qquad c_m \in \mathit{trap}(\varsigma,[\![\mathsf{apply}]\!]) \\ \mathit{text}_{\mathrm{this}}=\mathit{evalSimple}(s_0,\rho,\sigma,\kappa) \qquad \mathit{text}_{\mathrm{arg}}=\mathit{evalSimple}(s_1,\rho,\sigma,\kappa) \\ \kappa_m=(\bot,c_m,\mathit{text}_{\mathrm{arg}},a_0,\sigma_m) \qquad \kappa_r=(\bot,\bot,\bot,a_0,\sigma_m) \qquad \sigma_m=\sigma_\varsigma\sigma_M \\ \mathit{evalCall}(c_m,\mathit{text}_{\mathrm{arg}},\sigma_m,\langle\rangle,\kappa_r,\Xi,\kappa_m)\mapsto^* \mathbf{ko}(\mathit{text}_r,\sigma_r,\langle\rangle,\kappa_r,_) \qquad \sigma'_M=\sigma_r|_{\mathscr{R}_{\sigma_r}(a_M)}\end{array}}{(\mathit{text}_r,\sigma'_M)\in \mathit{handle}(\varsigma,\sigma_M)}$$

$$\textsc{Handle-No-Intercept}\quad \frac{\text{no intercept for } \varsigma}{(\bot,\sigma_M)\in \mathit{handle}(\varsigma,\sigma_M)}$$

Rule Handle-Method-Call applies when state $\varsigma$ is effectively a method call, which is the case when transitioning from an **ev** state with a method call as control component. Relation *trap* is used to obtain the callable *apply* trap. The remainder of the rule specifies the arguments for the call to the semantic *evalCall* function, which actually invokes the trap. Like a regular function application in $\textsc{js}_0$, the trap function is called with an empty local continuation. The trap function is called with the empty meta-continuation as if its body were top-level code. As a consequence, upon return of the trap function there is no continuation possible and a final state is reached.

Side effects can occur during the abstract interpretation of the handler trap in $\varsigma$. Therefore, the EE maintains a meta store ($\sigma_M$) and propagates those changes thorough the exploration of the base program's flow graph states. Before the *handle* apply call, the rules computes a new store $\sigma_m$ for $\varsigma$ where the information of the meta store $\sigma_M$ is "merged" with the $\varsigma$ store. After the abstract interpretation a new meta store $\sigma'_M$ is computed containing the possible side effects that happened during call.

Rule Handle-No-Intercept applies when no intercept is required for a state, i.e., when no other rules for *handle* apply. It returns an absent trap invocation result $\bot$ and the unmodified meta store.

Function $\mathscr{R} : \mathscr{P}(\mathit{Addr}) \times \mathscr{P}(\mathit{Addr}) \times \mathit{Store} \to \mathscr{P}(\mathit{Addr})$ computes the set of all addresses that are reachable from a given root set of addresses. In general terms, the overloaded function $\mathscr{T} : X \to \mathscr{P}(\mathit{Addr})$ returns the set of addresses directly referenced by components in the state space.

$$\begin{aligned}\mathscr{T}((f,\rho)) &= \mathscr{T}(\rho)\\ \mathscr{T}(a) &= \{a\}\\ \mathscr{T}(\omega) &= \mathscr{T}(\mathrm{Range}(\omega))\\ \mathscr{T}(\rho) &= \mathrm{Range}(\rho)\\ \mathscr{T}(\delta) &= \varnothing\\ \mathscr{T}(\{x_0,\ldots,x_n\}) &= \bigcup_{i\in 0..n}\mathscr{T}(x_i)\end{aligned}$$

Reachable addresses

$$\frac{a' \in \mathscr{T}(\sigma(a))}{a \rightsquigarrow_{\mathscr{T},\sigma} a'} \qquad\qquad \frac{a \in \mathscr{T}(\mathit{text}) \qquad a \rightsquigarrow^*_{\mathscr{T},\sigma} a'}{a \in \mathscr{R}_\sigma(\mathit{text})}$$

$$\frac{\begin{array}{c}\textbf{ev}([\![s.v]\!],\ldots)=\varsigma \qquad c_m\in \mathit{trap}(\varsigma,\mathsf{get})\\ \mathit{text}_{\mathsf{this}}=\mathit{evalSimple}(s,\rho,\sigma,\kappa) \qquad \mathit{text}_v=\mathit{evalSimple}(v,\rho,\sigma,\kappa)\\ \kappa_m=(\bot,c_m,\bot,a_0,\sigma_m) \qquad \kappa_r=(\bot,\bot,\bot,a_0,\sigma_m) \qquad \sigma_m=\sigma_\varsigma\sigma_M\\ \mathit{evalCall}(c_m,[\mathit{text}_{\mathsf{this}},\mathit{text}_v],\sigma_m,\langle\rangle,\kappa_r,\Xi,\kappa_m)\mapsto^{*}\textbf{ko}(\mathit{text}_r,\sigma_r,\langle\rangle,\epsilon,_)\\ \sigma'_M=\sigma_r|_{\mathcal{R}_{\sigma_r}(a_M)}\end{array}}{(\mathit{text}_r,\sigma'_M)\in \mathit{handle}(\varsigma,\sigma_M)}\;\textsc{Handle-Property-Read}$$

$$\frac{\begin{array}{c}\textbf{ev}([\![s_0.v\texttt{=}s_1]\!],\ldots)=\varsigma \qquad c_m\in \mathit{trap}(\varsigma,\mathsf{set})\\ \mathit{text}_0=\mathit{evalSimple}(s_0,\rho,\sigma,\kappa) \qquad \mathit{text}_1=\mathit{evalSimple}(s_1,\rho,\sigma,\kappa)\\ \kappa_m=(\bot,c_m,\mathit{text}_0,a_0,\sigma_m) \qquad \kappa_r=(\bot,\bot,\bot,a_0,\sigma_m) \qquad \sigma_m=\sigma_\varsigma\sigma_M\\ \mathit{evalCall}(c_m,[\mathit{text}_0,v,\mathit{text}_1],\sigma_m,\langle\rangle,\epsilon,\Xi,\kappa_m)\mapsto^{*}\textbf{ko}(\mathit{text}_r,\sigma_r,\langle\rangle,\epsilon,_)\\ \sigma'_M=\sigma_r|_{\mathcal{R}_{\sigma_r}(a_M)}\end{array}}{(\mathit{text}_r,\sigma'_M)\in \mathit{handle}(\varsigma,\sigma_M)}\;\textsc{Handle-Property-Write}$$

$$\frac{\begin{array}{c}\textbf{ev}([\![\mathsf{new}\ v(s)]\!],\ldots)=\varsigma \qquad c_m\in \mathit{trap}(\varsigma,\mathsf{construct})\\ \mathit{text}_0=\mathit{evalSimple}(v,\rho,\sigma,\kappa) \qquad \mathit{text}_1=\mathit{evalSimple}(s,\rho,\sigma,\kappa)\\ \kappa_m=(\bot,c_m,\mathit{text}_0,a_0,\sigma_m) \qquad \kappa_r=(\bot,\bot,\bot,a_0,\sigma_m) \qquad \sigma_m=\sigma_\varsigma\sigma_M\\ \mathit{evalCall}(c_m,[\mathit{text}_0,\mathit{text}_1],\sigma_m,\langle\rangle,\kappa_r,\Xi,\kappa_m)\mapsto^{*}\textbf{ko}(\mathit{text}_r,\sigma_r,\langle\rangle,\epsilon,_)\\ \sigma'_M=\sigma_r|_{\mathcal{R}_{\sigma_r}(a_M)}\end{array}}{(\mathit{text}_r,\sigma'_M)\in \mathit{handle}(\varsigma,\sigma_M)}\;\textsc{Handle-New-Obj-Expressions}$$

$$\frac{\begin{array}{c}\textbf{ev}([\![v\texttt{=}e]\!],\ldots)=\varsigma \qquad c_m\in \mathit{trap}(\varsigma,\mathsf{write})\\ \mathit{text}_0=\mathit{evalSimple}(e,\rho,\sigma,\kappa) \qquad \mathit{text}_M\in \mathit{lookupProp}([\![\mathsf{META}]\!],a_0,\sigma_\varsigma)\\ \kappa_m=(\bot,c_m,\mathit{text}_{\mathsf{arg}},a_0,\sigma_m) \qquad \kappa_r=(\bot,\bot,\bot,a_0,\sigma_m) \qquad \sigma_m=\sigma_\varsigma\sigma_M\\ \mathit{evalCall}(c_m,[v,\mathit{text}_0],\sigma_m,\langle\rangle,\epsilon,\Xi,\kappa_m)\mapsto^{*}\textbf{ko}(\mathit{text}_r,\sigma_r,\langle\rangle,\epsilon,_) \qquad \sigma'_M=\sigma_r|_{\mathcal{R}_{\sigma_r}(a_M)}\end{array}}{(\mathit{text}_r,\sigma'_M)\in \mathit{handle}(\varsigma,\sigma_M)}\;\textsc{Handle-Var-Assignment-Expressions}$$

**Figure 11** Additional rules for trapping program operations during the second static analysis phase.

Figure 11 shows the other rules for trapping operations which are specified in a similar fashion. Rules HANDLE-PROPERTY-READ and HANDLE-PROPERTY-WRITE handle object property access and updates, respectively. Rule HANDLE-NEW-OBJ-EXPRESSIONS handles new Obj() expressionsm and rule HANDLE-VAR-ASSIGNMENT-EXPRESSIONS handles variable assignment.

### E.3 Execution Exploration while Maintaining Meta State

For stateless meta code it suffices to visit all explored states once in an unspecified order and passing them to relation *handle*, ignoring the meta store from that relation. If, for a particular state, the value returned by *handle* subsumes the abstracted META.HALT, then this indicates that, according to the static analysis, a base program operation was intercepted that should halt the execution.

In case the meta code is stateful, then the meta state has to be maintained as well. In this case, detecting traps that halt the execution of a program must be expressed as a fixed point computation over the flow graph. Let $(\hookrightarrow) \sqsupseteq D \times State \times Store \times D \times State \times Store$ be the relation that operates on triples representing a state and the result of "handling" a state through $handle$, i.e., a meta value and meta store. The single rule below describes a transition from a reachable triple for one state to another triple for its successor state based on an edge in the flow graph and the result of handling the successor state. The initial triple for $\hookrightarrow$ is the initial state $\varsigma_0$ of the flow graph and the result of $handle(state_0)$.

$$\frac{\varsigma \longrightarrow \varsigma' \in \mathscr{G}(e) \qquad (text'_r, \sigma'_M) \in handle(\varsigma', \sigma_M)}{(_, \varsigma, \sigma_M) \hookrightarrow (text'_r, \varsigma', \sigma'_M)} \text{EE-TRANS}$$

Computing the transitive closure of $\hookrightarrow$ then enables detecting states for which a trap returns META.HALT.

$$\frac{(text_0, \sigma_0) \in handle(\varsigma_0, []) \qquad (text_0, \varsigma_0, \sigma_0) \hookrightarrow^* (text_r, \varsigma, \sigma_M) \qquad \alpha([\![\text{META.HALT}]\!]) \sqsubseteq text_r}{halt(\varsigma)}$$

## About the authors

**Angel Luis Scull Pupo** Angel Luis Scull Pupo is a PhD Student at the Software Languages Lab (SOFT) of the Vrije Universiteit Brussel (VUB) in Belguim. He obtained his Master in Applied Mathematics and Informatics to Administration in 2015 from the Universidad de Holguín (Uho). You can contact him at angel.luis.scull.pupo@vub.be.

**Jens Nicolay** Jens Nicolay is a Professor at the Software Languages Lab (SOFT) of the Vrije Universiteit Brussel (VUB) in Belgium. His doctoral research and work as a professor has mainly focused on program analysis with security as an important application. You can contact him at jens.nicolay@vub.be.

**Elisa Gonzalez Boix** Elisa Gonzalez Boix is an Associate Professor at the Software Languages Lab (SOFT) of the Vrije Universiteit Brussel (VUB) in Belgium. She obtained her Master in Informatics Engineering in 2004 from the Universitat Politècnica de Catalunya (UPC) and her PhD in Sciences in 2012 from the VUB. Since 2014, she leads a group on concurrent and distributed systems, studying programming language technology and dynamic software tools like debuggers. You can contact her at egonzale@vub.be.